\documentclass[letterpaper,journal]{IEEEtran}
\usepackage{amsmath,amsfonts}
\usepackage{algorithmic}
\usepackage{array}
\usepackage[caption=false,font=normalsize,labelfont=sf,textfont=sf]{subfig}
\usepackage{textcomp}
\usepackage{stfloats}
\usepackage{url}
\usepackage{verbatim}
\usepackage{graphicx}
\usepackage{amsmath,amssymb}
\usepackage{cuted}

\def\BibTeX{{\rm B\kern-.05em{\sc i\kern-.025em b}\kern-.08em
    T\kern-.1667em\lower.7ex\hbox{E}\kern-.125emX}}
\usepackage{balance}

\begin{document}
\title{A Multi-view Low-latency Direction Sensing for Low-altitude Networks via a Multi-armed Array}
\author{Jinbing Jiang, Feng Shu,  Jiatong Bai, Minghao Chen, Maolin Li and Jiangzhou Wang
    \thanks{\textit{(Corresponding author: Feng Shu.)}}
	\thanks{Jinbing Jiang, Jiatong Bai, Minghao Chen and Maolin Li are with the School of Information and Communication Engineering, Hainan University, Haikou, 570228, China (e-mails: jiangjinbing1125@163.com; 18419229733@163.com; 996739@hainanu.edu.cn; limaolin0302@163.com ).}
	\thanks{Feng Shu is with the School of Information and Communication Engineering, Hainan University, Haikou, 570228, China, and also with the School of Electronic and Optical Engineering, Nanjing University of Science and Technology, Nanjing, 210094, China (e-mail: shufeng0101@163.com).}
	\thanks{Jiangzhou Wang is with the National Mobile Communications Research Laboratory, Southeast University, Nanjing 210096, China, and also with Purple Mountain Laboratories, Nanjing 210096, China (e-mail: j.z.wang@seu.edu.cn).}}


\maketitle
\fontsize{10pt}{12pt}\selectfont
\begin{abstract}

Due to the directional antenna patterns, the received signal power will be severely attenuated when the target deviates significantly from the array boresight, resulting in a dramatic degradation in sensing accuracy in low-altitude scenarios with a fixed array at a base station. Conventional rotatable arrays have been adopted to address the low-altitude coverage problem via an iterative rotation process. However, this results in high sensing latency. In this paper, a multi-armed array is proposed to achieve low-latency low-altitude coverage with the ability to mitigate directional gain loss. In this new-type multiple-input multiple-output (MIMO) architecture, a large MIMO array consists of multiple subarrays with each subarray covering  different three-dimensional spatial sectors, enabling multi-view direction sensing for both low-altitude and terrestrial areas without array rotation. Then, a general fusion framework with two fusion methods, called signal-to-noise ratio (SNR) and Logistic-form (Log), is developed. The SNR fusion factors are proportional to the estimated SNR of the corresponding subarray, whereas the Log method uses an exponential function of SNR in dB to define the fusion factors. Finally, the Cramer-Rao lower bound  is derived to evaluate the sensing performance of the proposed array. Simulation results show that the proposed fusion methods perform much better than equal-gain fusion while the Log fusion may achieve better sensing accuracy in the low-SNR region. The proposed new array makes a significant reduction in sensing latency compared with conventional rotatable arrays with good sensing performance.

\end{abstract}


\section{Introduction}
\IEEEPARstart{W}{ith} the rapid development of sixth-generation (6G) low-altitude networks, the widespread application of unmanned aerial vehicles (UAVs) has placed increasing demands on real-time sensing and reliable wireless communications. Direction-of-arrival (DOA) estimation provides essential directional information for target localization, beamforming, and directional modulation \cite{Shu_Secure,Li2026_UAV_DM}. In particular, high-precision and low-latency DOA sensing is indispensable for integrated sensing and communication (ISAC) in low-altitude networks, where accurate and timely UAV localization and tracking are required \cite{Shu2025_H2AD_ML}.

To improve DOA estimation accuracy, various methods have been proposed. 
Among them, the multiple signal classification (MUSIC) algorithm achieves high-resolution DOA estimation by exploiting the orthogonality between the signal and noise subspaces \cite{Schmidt_MUSIC}. To reduce the computational complexity of spectral searching, the Root-MUSIC algorithm was developed to estimate DOAs through polynomial root finding \cite{Barabell1983_RootMUSIC}. Based on these classical algorithms, several methods have been developed to improve DOA estimation performance, including DOA ambiguity resolution in hybrid analog-and-digital (HAD) multiple-input multiple-output (MIMO) receivers \cite{Shu2018_HAD_DOA} and virtual array aperture enhancement \cite{Lai2025_TVT_DOA}. In addition, movable antenna and fluid antenna systems have been investigated to improve DOA estimation performance by exploiting flexible antenna positioning \cite{Ma2024_MA_DOA,Xu2026_FAS_DOA}.

Conventional antenna arrays at base stations (BSs) are typically designed for terrestrial communications, with their antenna boresights directed toward ground users \cite{Du2022_UAV_Uptilt}. Due to the directional antenna pattern, the antenna gain decreases as the target direction deviates from the array boresight \cite{Zheng_RA}. When sensing low-altitude targets, the antenna gain reduction leads to a lower received signal-to-noise ratio (SNR), thereby severely degrading DOA estimation accuracy. Electronic beam steering can adjust the array beam direction but remains limited by the directional antenna pattern \cite{Xi2021_WideAngleScan}.

To address these limitations, rotatable antenna arrays have been investigated to improve directional gain by adjusting their boresights toward the target direction \cite{Zheng_RA_Sensing}. Based on this architecture, rotatable array-based DOA sensing methods have been developed \cite{Jiang_IGSS}. In particular, the recursive rotation Root-MUSIC (RR-Root-MUSIC) algorithm was proposed to improve DOA estimation accuracy through iterative direction estimation and array rotation \cite{Jiang_RA}.  However, repeated mechanical rotations introduce additional sensing latency, making it difficult to meet the low-latency sensing requirements of low-altitude networks.

In this paper, a multi-armed array is proposed to achieve high-precision and low-latency DOA sensing in low-altitude networks.  The multi-armed structure is inspired by the multi-armed bandit (MAB) concept in reinforcement learning. The main contributions of this paper are summarized as follows:
\begin{enumerate}
	\item To mitigate the antenna gain loss caused by the deviation of targets from the array boresight and the delays caused by repeated rotations of conventional rotatable arrays, a multi-armed array is proposed. It is composed of multiple subarrays with different orientations, providing wide three-dimensional (3D) spatial coverage and reducing directional gain loss. This structure enables multi-view direction sensing without mechanical rotation, thereby avoiding rotation-induced sensing latency.

	\item To fully exploit multi-view sensing information, a general fusion framework incorporating SNR and Logistic-form (Log) methods is developed. The fusion factors of the SNR method are defined to be proportional to the estimated SNRs of the corresponding subarrays, whereas those of the Log are determined by an exponential function of the estimated SNRs in dB. Furthermore, the corresponding Cramer-Rao lower bound (CRLB) is derived to evaluate the sensing performance. Simulation results show that both fusion methods achieve higher sensing accuracy than equal-gain fusion, with the Log method providing further improvements in the low-SNR region. Moreover, compared with conventional rotatable arrays, the proposed array maintains good sensing performance while significantly reducing sensing latency without iterative rotation.

\end{enumerate}

The remainder of this paper is organized as follows. Section II presents the system model. Section III introduces the proposed multi-view fusion framework. Section IV derives the CRLB for the multi-armed array. Section V presents and discusses the simulation results. Finally, Section VI concludes the paper.

\section{System Model}

\begin{figure}
	\centering
	\includegraphics[width=0.75\linewidth]{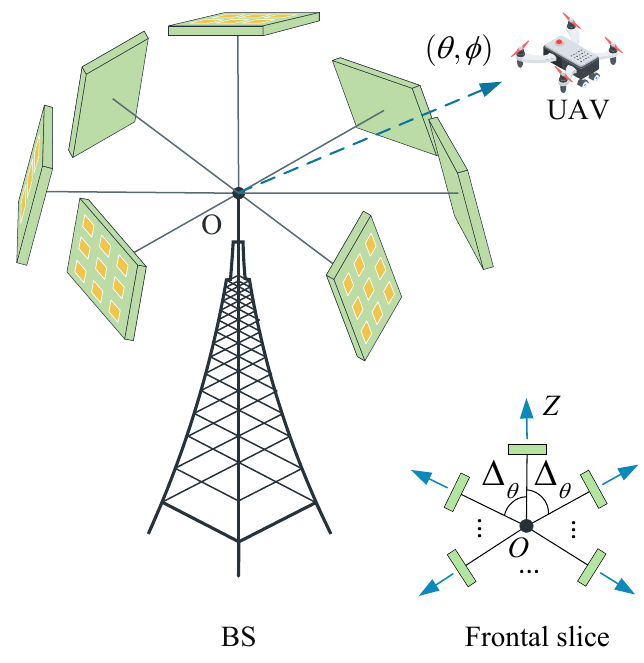}
	\caption{Proposed multi-armed distributed array for low-latency sensing and communication in low-altitude and terrestrial networks.}
	\label{fig:model}
\end{figure}

\subsection{Multi-armed Array}

As illustrated in Fig.~\ref{fig:model}, a multi-armed distributed antenna array is proposed to address the low-altitude coverage problem of traditional arrays. Such arrays may require rotation to achieve good low-altitude coverage and sensing, resulting in increased sensing latency. The proposed array significantly reduces the directional gain attenuation of antenna elements and provides wide coverage and multi-view sensing in low-altitude and terrestrial networks without rotating or moving the array.

A global Cartesian coordinate system (CCS) $o\text{-}xyz$ is established with its origin at the BS reference point. First, a set of subarrays is arranged in the global $x$-$z$ plane, with their boresights separated by an angular spacing $\Delta_\theta$. This arrangement is then extended around the global $z$-axis with an azimuthal spacing $\Delta_\phi$ to form a three-dimensional multi-armed array, excluding the negative $z$-axis direction. With one subarray oriented along the positive $z$-axis, the total number of subarrays is given by $Q=1+
\left(\left\lceil\frac{180^\circ}{\Delta_\theta} \right\rceil-1\right) \left\lceil\frac{360^\circ}{\Delta_\phi}\right\rceil$. Smaller angular spacings lead to denser subarray arrangements and higher sensing accuracy at a higher hardware cost, whereas larger ones result in sparser arrangements, lower sensing accuracy, and reduced hardware costs. Therefore, appropriate values of $\Delta_\theta$ and $\Delta_\phi$ can achieve a good balance between sensing performance and hardware cost.

Each subarray is composed of an $M\times N$ uniform planar array (UPA), with its center located at $\mathbf p_q=[p_{q,x},p_{q,y},p_{q,z}]^T$ in the global CCS, where $q=1,\ldots,Q$. A local CCS is defined at each subarray center, where the UPA lies in the $x'$-$z'$ plane and its boresight points along the positive $y'$-axis.

The boresight direction of the $q$-th subarray is defined by
\begin{align}
	\theta_q=i\Delta_\theta,\qquad
	\phi_q=j\Delta_\phi,
\end{align}
where $i=0,\ldots,\lceil180^\circ/\Delta_\theta\rceil-1$ and $j=0,\ldots,\lceil360^\circ/\Delta_\phi\rceil-1$, with $j=0$ for $i=0$.

The rotation matrix from the local CCS of subarray $q$ to the global CCS is given by
\begin{align}
	\mathbf R_q=
	\begin{bmatrix}
		\sin\phi_q & \sin\theta_q\cos\phi_q & -\cos\theta_q\cos\phi_q\\
		-\cos\phi_q & \sin\theta_q\sin\phi_q & -\cos\theta_q\sin\phi_q\\
		0 & \cos\theta_q & \sin\theta_q
	\end{bmatrix}.
\end{align}

Accordingly, the boresight unit vector of subarray $q$ in the global CCS is expressed by
\begin{align}
	\mathbf n_q
	= \mathbf R_q\mathbf e_{y'}= \begin{bmatrix}
		\sin\theta_q\cos\phi_q,~
		\sin\theta_q\sin\phi_q,~
		\cos\theta_q
	\end{bmatrix}^T,
\end{align}
where $\mathbf e_{y'}=[0,1,0]^T$ is the unit vector along the local $y'$-axis.

The position of the $(m,n)$-th antenna element relative to the subarray center in the local CCS is defined by $\bar{\mathbf r}_{m,n}=[x_m,0,z_n]^T$, where $x_m=(m-\frac{M-1}{2})d$ and $z_n=(n-\frac{N-1}{2})d$, with $m=0,\ldots,M-1$ and $n=0,\ldots,N-1$.  The antenna spacing is $d=\lambda/2$, where $\lambda$ is the carrier wavelength. Each subarray contains $N_s=MN$ antenna elements, resulting in a total of $N_{\mathrm{tot}}=QN_s$ elements at the BS.

The global position of the $(m,n)$-th antenna element in subarray $q$ is obtained as
\begin{align}
	\mathbf r_{q,m,n}
	=\mathbf p_q+\mathbf R_q\bar{\mathbf r}_{m,n}.
\end{align}

\subsection{Received Signal Model}

Consider a narrowband signal arriving from a single far-field target with direction $(\theta,\phi)$, where $\theta\in[0,\pi]$ and $\phi\in[-\pi,\pi)$ denote the elevation and azimuth angles, respectively. The corresponding unit direction vector is
\begin{equation}
	\mathbf u(\theta,\phi)
	=
	[\sin\theta\cos\phi,\,
	\sin\theta\sin\phi,\,
	\cos\theta]^T.
	\label{eq:pointing_vector}
\end{equation}

The spatial phase shift of the $(m,n)$-th antenna element relative to the center of subarray $q$ is calculated as
\begin{align}
	&\varphi_{q,m,n}(\theta,\phi)
	=\kappa\bar{\mathbf r}_{m,n}^T
	\mathbf R_q^T\mathbf u(\theta,\phi)=\kappa\Big[
	x_m\sin\theta\sin(\phi_q-\phi) \nonumber\\
	&\quad+z_n\big(
	-\cos\theta_q\sin\theta\cos(\phi-\phi_q)
	+\sin\theta_q\cos\theta
	\big)
	\Big].
	\label{eq:element_phase}
\end{align}
where $\kappa=2\pi/\lambda$. The corresponding antenna element response is
\begin{equation}
	b_{q,m,n}(\theta,\phi)
	=
	e^{j\varphi_{q,m,n}(\theta,\phi)}.
\end{equation}

The steering vector of subarray $q$ is constructed as
\begin{equation}
	\mathbf b_q(\theta,\phi)
	=
	[b_{q,0,0}(\theta,\phi),\ldots,
	b_{q,M-1,N-1}(\theta,\phi)]^T.
\end{equation}

Let $\zeta_q $ denote the deviation angle between the target direction and the boresight of subarray $q$, which is calculated as
\begin{align}
	\zeta_q&=\arccos\left(\mathbf n_q^T\mathbf u(\theta,\phi)\right)\nonumber\\
	&=\arccos\left(\sin\theta_q\sin\theta
	\cos(\phi-\phi_q)
	+\cos\theta_q\cos\theta\right).
\end{align}

Following the directional antenna model in
\cite{Zheng_RA}, the power gain of each antenna element is modeled as
\begin{equation}
	G_q(\theta,\phi)
	=
	\begin{cases}
		G_0\cos^{2p}(\zeta_q),
		& 0\leq\zeta_q<\pi/2,\\
		0, & \mathrm{otherwise},
	\end{cases}
\end{equation}
where $p \geq 0$ is the directivity factor and $G_0=2(2p+1)$ is the maximum boresight gain. With path loss normalized, the corresponding amplitude gain is $g_q(\theta,\phi)=\sqrt{G_q(\theta,\phi)}$.

The center position of subarray $q$ introduces a common phase factor
\begin{equation}
	\chi_q(\theta,\phi)
	=e^{j\kappa\mathbf p_q^T\mathbf u(\theta,\phi)}.
\end{equation}

The received signal vector of subarray $q$ at snapshot $\tau$ is modeled as
\begin{equation}
	\mathbf y_q(\tau)
	=
	\chi_q(\theta,\phi)
	g_q(\theta,\phi)
	\mathbf b_q(\theta,\phi)s(\tau)
	+\mathbf w_q(\tau),
	\label{eq:received_signal_subarray}
\end{equation}
where $s(\tau)$ is the transmitted narrowband signal with  $\sigma_s^2=\frac{1}{T}\sum_{\tau=1}^{T}|s(\tau)|^2$, and $\mathbf w_q(\tau)\sim \mathcal{CN} (\mathbf 0, \sigma^2\mathbf I_{N_s})$ is the additive white Gaussian noise (AWGN) vector at subarray $q$.

\section{Proposed Multi-view Fusion Framework}

\begin{figure*}
	\centering
	\includegraphics[width=0.75\linewidth]{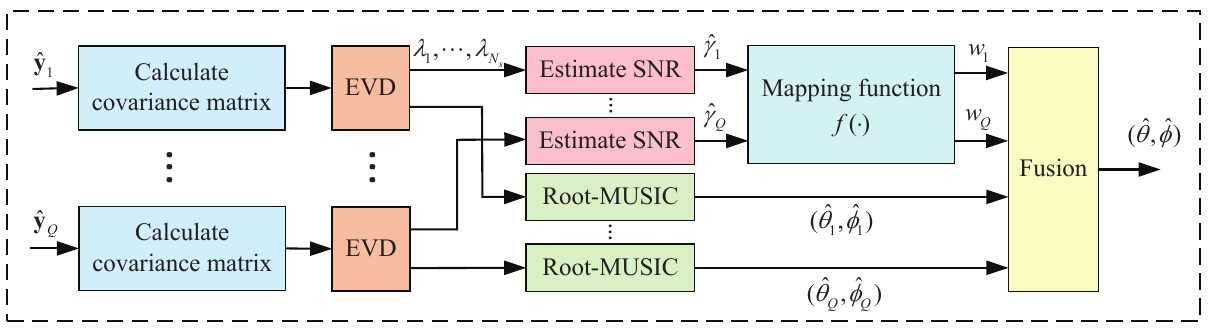}
	\caption{Proposed multi-view fusion framework. }
	\label{fig:flowchart}
\end{figure*}

In this section, a general multi-view DOA fusion framework is developed to improve direction-sensing accuracy by combining the DOA estimates from different subarrays. As illustrated in Fig.~\ref{fig:flowchart}, the proposed framework includes two fusion methods, namely SNR fusion and Log fusion.

\subsection{Proposed SNR Fusion}

For subarray $q$, the received signal matrix with $T$ snapshots is defined as
\begin{equation}
	\mathbf Y_q
	=\left[\mathbf y_q(1),\ldots,\mathbf y_q(T)\right].
\end{equation}

The sample covariance matrix is calculated as
\begin{equation}
	\hat{\mathbf C}_q
	=\frac{1}{T}\mathbf Y_q\mathbf Y_q^H.
	\label{eq:sample_covariance}
\end{equation}

The eigenvalue decomposition (EVD) of
$\hat{\mathbf C}_q$ yields
\begin{equation}
	\hat{\mathbf C}_q
	=\mathbf V_q\mathbf\Lambda_q\mathbf V_q^H,
	\label{eq:covariance_evd}
\end{equation}
where $\mathbf V_q$ contains the orthonormal eigenvectors and $\mathbf\Lambda_q =\operatorname{diag}(\lambda_{q,1},\ldots,\lambda_{q,N_s})$, with $\lambda_{q,1}\geq\cdots\geq\lambda_{q,N_s}$.

Based on \eqref{eq:covariance_evd}, the two-dimensional Root-MUSIC algorithm in \cite{Jiang_RA} is applied to estimate the local DOA unit vector $\hat{\mathbf u}_q^{\mathrm{loc}}$, which is  transformed into the global CCS as
$\hat{\mathbf u}_q=\mathbf R_q\hat{\mathbf u}_q^{\mathrm{loc}}$,
where $\hat{\mathbf u}_q=
[\hat u_{q,x},\hat u_{q,y},\hat u_{q,z}]^T$. The corresponding elevation and azimuth angle estimates are obtained as
\begin{align}
	\hat{\theta}_q=\arccos\!\left(\hat u_{q,z}\right),\quad
	\hat{\phi}_q=\operatorname{atan2}\!\left(\hat u_{q,y}, \hat u_{q,x}\right).
\end{align}

The noise power of subarray $q$ is estimated using the eigenvalues as
\begin{equation}
	\hat{\sigma}_q^2
	=
	\frac{1}{N_s-K}
	\sum_{i=K+1}^{N_s}\lambda_{q,i},
	\label{eq:noise_power_estimation}
\end{equation}
where $K=1$ is the number of signal sources. The received SNR is estimated as
\begin{equation}
	\hat{\gamma}_q
	=
	\max\left\{
	\frac{
		\sum_{i=1}^{N_s}\lambda_{q,i}
		-
		N_s\hat{\sigma}_q^2
	}{
		N_s\hat{\sigma}_q^2
	},
	0
	\right\}.
	\label{eq:snr_estimation}
\end{equation}

Let $\mathcal I_S$ denote the index set of the $S$ subarrays with the highest estimated SNRs. The normalized fusion weight for subarray $q$ is defined as
\begin{equation}
	w_q^{\mathrm{SNR}}
	=
	\frac{\hat{\gamma}_q}
	{\sum_{j\in\mathcal I_S}\hat{\gamma}_j},
	\quad q\in\mathcal I_S,
	\label{eq:snr_weight}
\end{equation}
where $w_q^{\mathrm{SNR}}=0$ for $q\notin\mathcal I_S$, and
$\sum_{q=1}^{Q}w_q^{\mathrm{SNR}}=1$.

The fused elevation angle is obtained by
\begin{equation}
	\hat{\theta}_{\mathrm{SNR}}
	=\sum_{q=1}^{Q}
	w_q^{\mathrm{SNR}}\hat{\theta}_q.
	\label{eq:snr_polar_fusion}
\end{equation}

For azimuth sensing, a weighted circular mean is adopted to account for angular periodicity, yielding
\begin{equation}
	\hat{\phi}_{\mathrm{SNR}}
	=\operatorname{atan2}\!\left(
	\sum_{q=1}^{Q}w_q^{\mathrm{SNR}}\sin\hat{\phi}_q,\,
	\sum_{q=1}^{Q}w_q^{\mathrm{SNR}}\cos\hat{\phi}_q
	\right).
	\label{eq:snr_azimuth_fusion}
\end{equation}

\subsection{Proposed Logistic-Form Fusion}

Inspired by the MAB framework in reinforcement learning,  the proposed Log fusion method employs an exponential weighting function with an adjustable base to control the contributions of selected subarrays. Specifically, the estimated SNR of each selected subarray \(q\) is converted to decibels as
\begin{equation}
	\hat{\gamma}_{q,\mathrm{dB}}
	=
	10\log_{10}\hat{\gamma}_q, \quad q\in\mathcal I_S.
	\label{eq:snr_db}
\end{equation}

The normalized fusion weight of subarray $q$ is defined as
\begin{equation}
	w_q^{\mathrm{Log}}
	=
	\frac{\rho ^{\hat{\gamma}_{q,\mathrm{dB}}}}
	{\sum_{j\in\mathcal I_S}\rho ^{\hat{\gamma}_{j,\mathrm{dB}}}},\quad q\in\mathcal I_S,
	\label{eq:logistic_weight}
\end{equation}
where $\rho >1$ is the adjustable base. For $q\notin\mathcal I_S$, $w_q^{\mathrm{Log}}=0$,
and $\sum_{q=1}^{Q}w_q^{\mathrm{Log}}=1$.

As $\rho\rightarrow 1^+$, the weights approach $1/S$, corresponding to equal-gain fusion. When $\rho=10^{1/10}$, Log fusion becomes
equivalent to SNR fusion. For $\rho>10^{1/10}$, greater weights are assigned to subarrays with higher estimated SNRs.

The final DOA estimates are given by
\begin{equation}
	\hat{\theta}_{\mathrm{Log}}
	=
	\sum_{q=1}^{Q}
	w_q^{\mathrm{Log}}\hat{\theta}_q,
	\label{eq:log_polar_fusion}
\end{equation}
\begin{equation}
	\hat{\phi}_{\mathrm{Log}}
	=
	\operatorname{atan2}\!\left(
	\sum_{q=1}^{Q}w_q^{\mathrm{Log}}\sin\hat{\phi}_q,\,
	\sum_{q=1}^{Q}w_q^{\mathrm{Log}}\cos\hat{\phi}_q
	\right).
	\label{eq:log_azimuth_fusion}
\end{equation}

\subsection{Delay Analysis and Comparison}


Multiple iterations of array mechanical rotation are required for the rotatable array, in which each rotation is achieved by stepper motor and will last seconds \cite{Xie2026}. In contrast, the proposed multi-armed array eliminates mechanical rotation, reducing the sensing latency to the microsecond scale. Therefore, the designed array achieves significant advantage in sensing latency, compared with conventional rotation-based methods.

\section{Derivation of the CRLB}

In this section, the CRLB of the proposed multi-armed array is derived to evaluate its theoretical direction-sensing performance.

The unknown parameter vector is defined as $\boldsymbol{\psi}=[\theta,\phi]^T$, and the transmitted signal $s(\tau)$ is assumed to be known. Since each subarray is processed independently, its common phase is treated as an unknown nuisance parameter. Owing to the centered UPA geometry, the corresponding Fisher information matrix (FIM) cross terms with $\theta$ and $\phi$ vanish, allowing the common phase to be eliminated without affecting the DOA FIM. Accordingly, the received signal at the $(m,n)$-th antenna element of subarray $q$ is expressed as
\begin{equation}
	y_{q,m,n}(\tau)
	=
	s(\tau)
	g_q(\boldsymbol{\psi})
	e^{j\varphi_{q,m,n}(\boldsymbol{\psi})}
	+
	w_{q,m,n}(\tau),
	\label{eq:crlb_signal}
\end{equation}
where $w_{q,m,n}(\tau)\sim\mathcal{CN}(0,\sigma^2)$ denotes complex Gaussian noise.

The mean of the received signal is expressed as
\begin{equation}
	\mu_{q,m,n}(\tau)
	=
	s(\tau)g_q(\boldsymbol{\psi})
	e^{j\varphi_{q,m,n}(\boldsymbol{\psi})}.
\end{equation}

Accordingly, the $(i,j)$-th element of the FIM for subarray $q$ is given by
\begin{equation}
	[\mathbf J_q]_{ij}
	=
	\frac{2}{\sigma^2}
	\Re\left\{
	\sum_{\tau=1}^{T}\sum_{m,n}
	\left(
	\frac{\partial\mu_{q,m,n}(\tau)}
	{\partial\psi_i}
	\right)^*
	\frac{\partial\mu_{q,m,n}(\tau)}
	{\partial\psi_j}
	\right\},
\end{equation}
where $\psi_i,\psi_j\in\{\theta,\phi\}$.

For $\cos\zeta_q>0$, the FIM elements of subarray $q$ are obtained as
{\footnotesize
	\begin{align}
		J_{\theta\theta}^{(q)}
		=&\frac{T\eta\pi^2}{12}
		\Big\{
		\nu c_q^{2p-2}a_q^2
		+2c_q^{2p}
		\big[
		\xi_x\cos^2\theta\sin^2\Delta\phi_q
		+\xi_zd_q^2
		\big]
		\Big\},
		\label{eq:Jtt}
		\\
		J_{\theta\phi}^{(q)}
		=&\frac{T\eta\pi^2}{12}
		\Big\{
		-\nu c_q^{2p-2}a_qb_q
		+2c_q^{2p}
		\big[
		\xi_x\sin\theta\cos\theta
		\sin\Delta\phi_q\cos\Delta\phi_q
		\nonumber\\
		&\quad
		-\xi_z\cos\theta_q\sin\theta
		\sin\Delta\phi_q d_q
		\big]
		\Big\},
		\label{eq:Jtp}
		\\
		J_{\phi\phi}^{(q)}
		=&\frac{T\eta\pi^2}{12}
		\Big\{
		\nu c_q^{2p-2}b_q^2
		+2c_q^{2p}
		\big[
		\xi_x\sin^2\theta\cos^2\Delta\phi_q
		\nonumber\\
		&\quad
		+\xi_z\cos^2\theta_q\sin^2\theta
		\sin^2\Delta\phi_q
		\big]
		\Big\},
		\label{eq:Jpp}
	\end{align}
}
where 
\begin{equation}
		\begin{aligned}
			a_q &=
			\sin\theta_q\cos\theta\cos\Delta\phi_q
			-\cos\theta_q\sin\theta,\\
			b_q &=
			\sin\theta_q\sin\theta\sin\Delta\phi_q,\\
			c_q &=
			\sin\theta_q\sin\theta\cos\Delta\phi_q
			+\cos\theta_q\cos\theta,\\
			d_q &=
			\cos\theta_q\cos\theta\cos\Delta\phi_q
			+\sin\theta_q\sin\theta.
		\end{aligned}
	\label{eq:coefficients}
\end{equation}
and $\eta=\frac{MN\sigma_s^2G_0}{\sigma^2}$,
	$\nu=\frac{24p^2}{\pi^2}$, $\xi_x=M^2-1$, $\xi_z=N^2-1$, and $\Delta\phi_q=\phi-\phi_q$.

For $\cos\zeta_q\leq0$, the directional antenna gain vanishes, resulting in $\mathbf J_q=\mathbf 0$. Since the noise vectors at different subarrays are independent, the overall FIM is obtained by summing the individual subarray FIMs as
\begin{equation}
	\mathbf J(\boldsymbol{\psi})
	=
	\sum_{q=1}^{Q}\mathbf J_q
	=
	\begin{bmatrix}
		J_{\theta\theta} & J_{\theta\phi}\\
		J_{\theta\phi} & J_{\phi\phi}
	\end{bmatrix},
\end{equation}
where $J_{ij}=\sum_{q=1}^{Q}J_{ij}^{(q)}$, $i,j\in\{\theta,\phi\}$.

The corresponding CRLBs for $\theta$ and $\phi$ are expressed as
\begin{align}
	\mathrm{CRLB}_{\theta}
	=
	\frac{
		J_{\phi\phi}
	}
	{
		J_{\theta\theta}J_{\phi\phi}
		-
		J_{\theta\phi}^{2}
	},
	\mathrm{CRLB}_{\phi}
	=
	\frac{
		J_{\theta\theta}
	}
	{
		J_{\theta\theta}J_{\phi\phi}
		-
		J_{\theta\phi}^{2}
	}.
	\end{align}

\section{Simulation Results and discussions}
In this section, simulation results are presented to evaluate the proposed SNR and Log methods, with the CRLB serving as a performance benchmark.  Simulation parameters are set as follows:  $M=N=4$, $T=100$, $p=2$, $\theta=5^\circ$, $\phi=55^\circ$, $\Delta=\Delta_\theta=\Delta_\phi=60^o$, $S=4$,  $\mathrm{SNR}=5$ dB, $\rho=1.5$. Here, SNR denotes the nominal SNR before accounting for the directional antenna gain. The elevation angle root mean square error (RMSE) is evaluated over 5000 Monte Carlo trials.

\begin{figure*}[htbp]
	\centering
	
	\begin{minipage}[t]{0.31\textwidth}
		\centering
		\includegraphics[width=\linewidth]{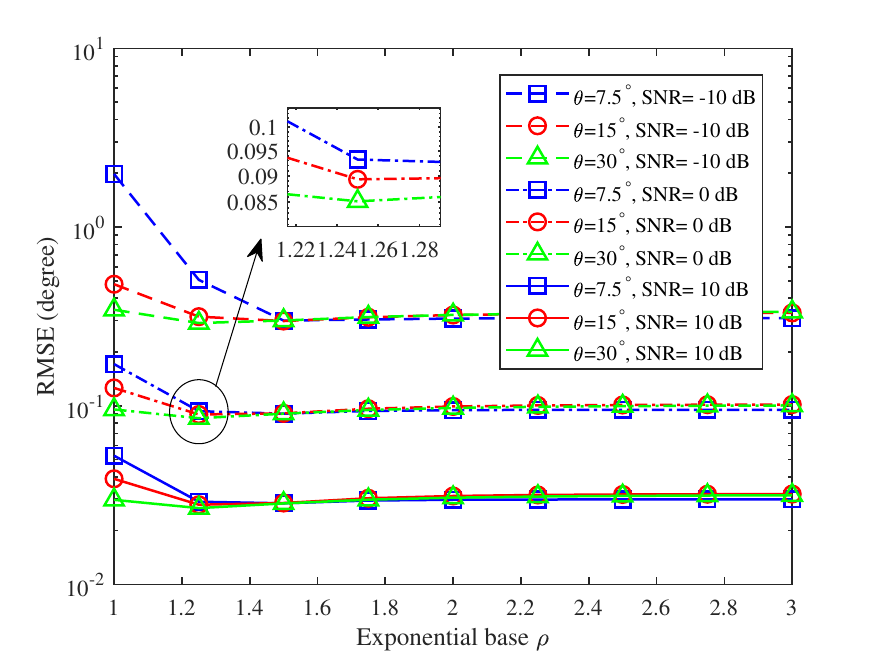} 
		\caption{RMSE versus exponential base $\rho$ for different SNRs and elevation angles $\theta$.}
		\label{fig:rho}
	\end{minipage}
	\hfill
	\begin{minipage}[t]{0.31\textwidth}
		\centering
		\includegraphics[width=\linewidth]{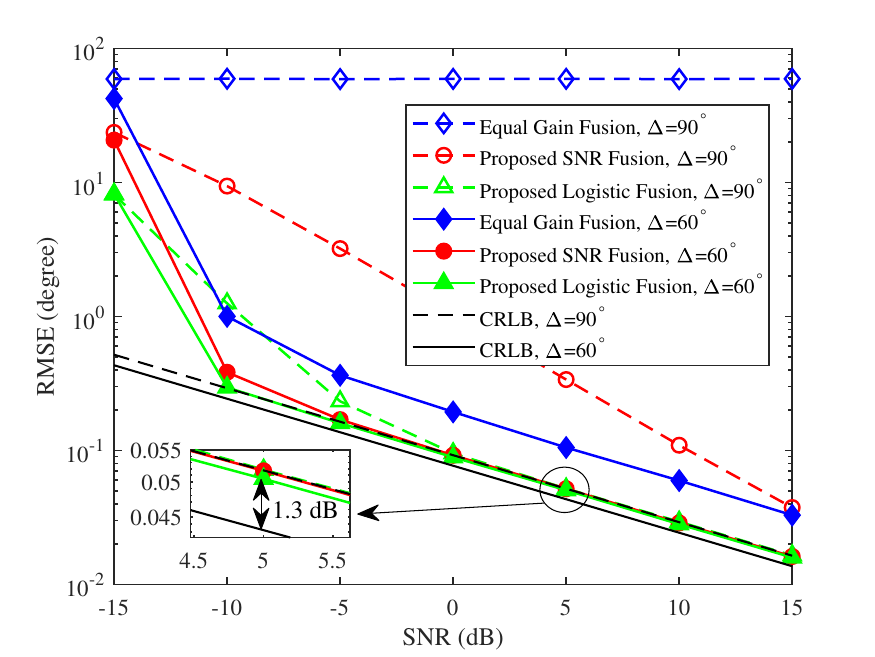} 
		\caption{RMSE versus SNR for different fusion methods and angular spacings $\Delta$.}
		\label{fig:SNR}
	\end{minipage}
	\hfill
	\begin{minipage}[t]{0.31\textwidth}
		\centering
		\includegraphics[width=\linewidth]{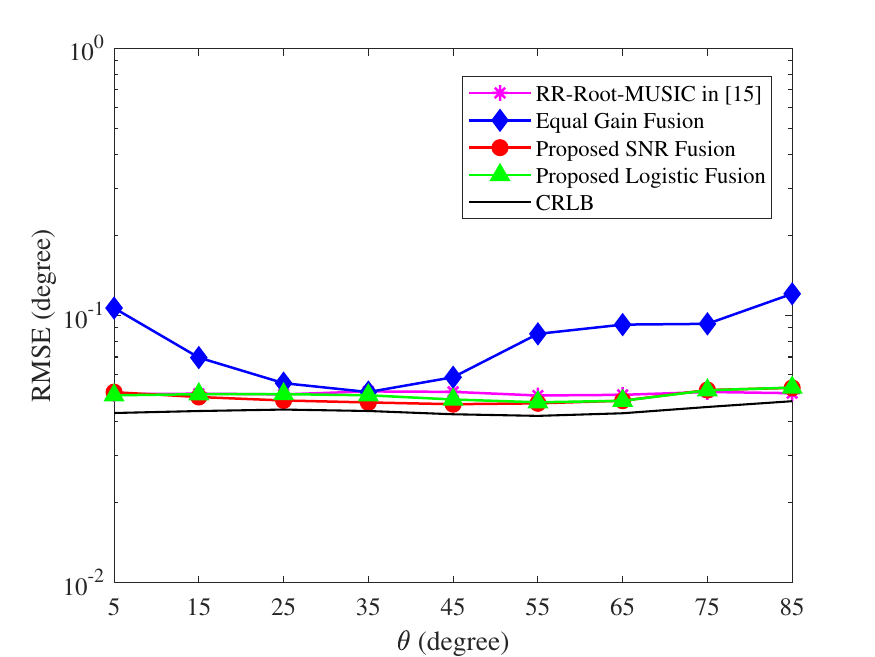} 
		\caption{RMSE versus elevation angle $\theta $ for different methods.}
		\label{fig:thetacompare60rr}
	\end{minipage}
\end{figure*}

Fig. \ref{fig:rho} illustrates the RMSE versus the exponential base $\rho$ for different elevation angles $\theta$ and SNRs. As observed, the RMSE curves exhibit similar trends under different conditions, decreasing significantly as $\rho$ increases from 1 and generally reaching their minima around $\rho=1.25$--$1.5$. Beyond this range, further increases in $\rho$ provide little improvement and may slightly degrade the sensing performance.

Fig. \ref{fig:SNR} shows the RMSE versus SNR for different methods with $\Delta=60^\circ$ ($S=4$) and $\Delta=90^\circ$ ($S=3$). It follows that both proposed methods significantly outperform equal-gain fusion, especially in the low-SNR region, and tightly approach the CRLB. Specifically, Log fusion with $\Delta=60^\circ$ incurs an SNR loss of approximately 1.3 dB relative to the CRLB. Furthermore, reducing $\Delta$ improves sensing accuracy across all fusion methods by providing more densely distributed sensing views. This improvement comes at the cost of more subarrays and higher hardware complexity, highlighting the tradeoff between sensing accuracy and hardware cost.

Fig. \ref{fig:thetacompare60rr} plots the RMSE versus elevation angle $\theta$ for different methods. RR-Root-MUSIC uses the same UPA size and number of snapshots per iteration. Equal-gain fusion exhibits noticeable performance degradation at certain angles because it assigns identical weights to subarrays with different received SNRs. The proposed SNR and Log fusion methods maintain stable sensing accuracy over the entire angular range from $5^\circ$ to $85^\circ$, with RMSEs remaining close to the CRLB. Both methods achieve accuracy comparable to RR-Root-MUSIC in \cite{Jiang_RA} without mechanical rotation or its associated delay.

\section{Conclusion}
In this paper, a novel multi-armed array has been proposed for low-latency and high-accuracy multi-view direction sensing in low-altitude networks. The proposed array utilizes subarrays with different orientations to enable multi-view sensing, thereby providing wide sensing coverage for the BS and reducing the sensing latency caused by repeated mechanical rotations in conventional rotatable arrays. Furthermore, a general fusion framework incorporating SNR and Log methods has been developed to integrate the DOA estimates across subarrays, and the CRLB of the proposed array has been derived to evaluate the sensing performance. Simulation results show that the Log fusion is more effective at very low SNRs, and both fusion methods exhibit superiority over equal-gain fusion. Crucially, compared with RR-Root-MUSIC, the proposed array achieves comparable DOA estimation accuracy while fundamentally avoiding rotation-induced sensing latency. These results highlight the potential of the proposed multi-armed array for low-latency direction sensing in future 6G low-altitude networks.

\bibliographystyle{IEEEtran}
\bibliography{refs}

\end{document}